\documentclass[superscriptaddress,onecolumn,amsmath,amssymb]{revtex4}
\usepackage{graphicx}
\usepackage{color}
\usepackage{ulem}

\begin{document}

\title{Equivalence of dissipative and dissipationless dynamics of interacting quantum systems with its application to the unitary Fermi gas}

\author{M. Tokieda}
\email[Present address: ]{Inria, Paris, France}
\email[E-mail address: ]{masaaki.tokieda@inria.fr}
\affiliation{Department of Physics, Tohoku University, Sendai, Japan}
\affiliation{Strangeness Nuclear Physics Laboratory, RIKEN Nishina Center, Wako, Japan}
\author{S. Endo}
\email[E-mail address: ]{shimpei.endo@nucl.phys.tohoku.ac.jp}
\affiliation{Department of Physics, Tohoku University, Sendai, Japan}
\affiliation{Frontier Research Institute for Interdisciplinary Science, Tohoku University, Sendai, Japan}

\begin{abstract}
  We analytically study quantum dissipative dynamics described by the Caldirola-Kanai model with inter-particle interactions. We have found that the dissipative quantum dynamics of the Caldirola-Kanai model can be exactly mapped to a dissipationless quantum dynamics under a negative external harmonic potential, even when the particles are strongly interacting. In particular, we show that the mapping is valid for the unitary Fermi gas, which is relevant for cold atoms and nuclear matters.
\end{abstract}

\maketitle


\section{Introduction}


Dissipation plays essential roles in the non-equilibrium dynamics of quantum matters. There has been rapid growth of research interests in the dissipative quantum dynamics as it is relevant to 
macroscopic quantum tunneling~\cite{CaldeiraLeggett81}, low-energy nuclear reactions~\cite{Diaztorres08,Frontiers07,Jeung21}, dynamics of cold atoms~\cite{diehl2008quantum,bardon2014transverse,cao2011universal}, and quantum information processings~\cite{Azouit17,RevModPhys.93}. Milestones in this research field is the Caldeira-Leggett model~\cite{CaldeiraLeggett81} and the Lindblad type equations~\cite{GKS76,Lindblad76}, where dissipation 
in the quantum system originates 
from the coupling of the system with the environment.



While these approaches to the quantum dissipative dynamics have been widely and successfully used, it is rather difficult to obtain an exact result in an analytical manner for most practical problems. 
To deal with the quantum dissipative dynamics in a more analytically feasible manner, we can alternatively resort to a simpler quantum equation of motion. One of the elementary models describing such a dissipative quantum motion is the Caldirola-Kanai model~\cite{Caldirola41,Kanai48} characterized by the Lagrangian
\begin{equation}
  L_{\rm CK}( \{ {\bf x}_i \}, \{ \dot{\bf {x}}_i \}, t) = e^{\gamma t} \left[ \sum_{i=1}^N \frac{m_i}{2} \dot{\bf {x}}_i^2 - U({\bf x}_1, \dots, {\bf x}_N,t) \right],
  \label{eq:L_CK}
\end{equation}
and its semi-classical equation of motion
\begin{equation}
\label{eq:Kanai}  \ddot{\bf x}_i(t) + \gamma \dot{\bf x}_i(t) + \frac{1}{m_i} \frac{\partial U}{\partial {\bf x}_i} ({\bf x}_1(t), \dots, {\bf x}_N(t),t) = 0.
\end{equation}
where $m_i$, ${\bf x}_i$ and $\dot{\bf {x}}_i = (d/dt) {\bf x}_i$ are the mass, the position, and the velocity of the $i$-th particle respectively. $U$ denotes an interaction or an external potential, and $\gamma$ is the dissipation rate (see Eq.~(\ref{eq:x_scheq}) for the Schr\"odinger equation and the Hamiltonian for the Caldirola-Kanai model).
Taking advantage of its simplicity, the Caldirola-Kanai model has been applied to various dissipative quantum phenomena, including damping of electromagnetic fields in a plasma medium~\cite{CLMM14}, dissipative quantum tunneling in low-energy nuclear fusion reactions~\cite{PhysRevC.95.054604}, 
dynamics of a damped charged oscillator in the presence of the Aharanov-Bohm effect~\cite{NACG21}, and so on.
While the Caldirola-Kanai model has usually been applied to a single-particle motion regarding $U$ as an external potential as in the original setting~\cite{Caldirola41,Kanai48}, 
we can also consider the dynamics of interacting quantum particles under the dissipation regarding $U$ as inter-particle interactions.

In this paper, we analytically study the quantum dissipative motion of interacting particles with the Caldirola-Kanai model.
We show that this dissipative system can be exactly mapped to a dissipationless system with an inverted harmonic potential. Similar mappings have been discussed for single-particle systems in the literature \cite{Kerner58,BJM94,HuangWu98}.
In this paper, we extend the mapping to many-particle systems, particularly to a strongly interacting system with a divergently large $s$-wave scattering length, i.e. unitary interaction.
Such a system has attracted increasing interests in terms of the unitary Fermi gas in cold atoms. Furthermore, as neutron matters are well described by the unitary Fermi gas~\cite{bertsch2001many,zwerger2011bcs,RevModPhys.80.1215,horikoshi2019cold,PhysRevA.97.013601,endovirialsp,PhysRevC.82.045802}, our exact mapping should be useful for understanding non-equilibrium dynamics of nuclear matters.

The paper is organized as follows: in Sec.~\ref{sec:map_nonint}, we introduce the Caldirola-Kanai model and its Hamiltonian, and then show the formal mapping procedure from the Caldirola-Kanai model to a dissipationless Hamiltonian with an inverted harmonic barrier. In Sec.~\ref{sec:map_int}, we study the exact mapping for interacting systems. In particular, we consider the Caldirola-Kanai model with the zero-range interaction, and show that it can be mapped to the dissipationless Hamiltonian with an inverted harmonic barrier when the particles are interacting via the unitary interaction.
We conclude and discuss physical implications of our exact mapping in Sec.~\ref{sec:concl}.

\section{Mapping the Caldirola-Kanai model to a dissipationless motion with an inverted harmonic barrier}\label{sec:map_nonint}

In this section, we show the exact mapping from the Caldirola-Kanai model to a dissipationless Hamiltonian with an inverted harmonic potential. To make it self-contained, we reformulate the arguments in Refs.~\cite{Kerner58,BJM94,HuangWu98} and begin our discussion with a single-particle system in a one-dimensional space for classical mechanics in Sec.~\ref{sec:Nonint_cl} and for quantum mechanics in Sec.~\ref{sec:Nonint_qm}. We see that the mapping between the dissipative and dissipationless dynamics can be understood as a transformation of the coordinate from one system to the other. We then extend this idea to a many-particle system in a $d$-dimensional space in Sec.~\ref{sec:Nonint_mp}. We find that the mapping can formally remain true even in the presence of 
inter-particle interactions.


\subsection{\label{sec:Nonint_cl}Transformation of the classical coordinates}

To see the mapping heuristically, let us first consider the classical dynamics of a damped harmonic oscillator in one spacial dimension.
The equation of motion reads
\begin{equation}
  \ddot{x}(t) + \gamma \dot{x}(t) + \omega^2 x(t) = 0,
  \label{eq:DHO_eom}
\end{equation}
with the frequency $\omega$ and the dissipation rate $\gamma$. Its analytical solution can be obtained easily. For example, for an under-damped case, it reads
\begin{equation}
  x(t) = e^{-\frac{\gamma t }{2}} \left[x(0) \cos(\Omega t) + \frac{\dot{x}(0) + (\gamma/2)x(0)}{\Omega} \sin(\Omega t) \right],
  \label{eq:DHO_solution}
\end{equation}
with $\Omega = \sqrt{\omega^2 - (\gamma/2)^2}$.

Notice that the solution is similar to that of a simple harmonic oscillator.
This similarity can be understood in the following way.
We first introduce a new variable $y(t)$,
\begin{equation}
  x(t) = e^{- \gamma t /2} y(t).
  \label{eq:x_vs_y}
\end{equation}
Substituting this relation into the equation of motion for $x(t)$, Eq.~(\ref{eq:DHO_eom}), one finds that $y(t)$ satisfies the equation of motion of a simple harmonic oscillator,
\begin{equation}
  \ddot{y}(t) + \Omega^2 y(t) = 0.
  \label{eq:HO_eom}
\end{equation}
The general solution is given by
\begin{equation}
  y(t) = y(0) \cos(\Omega t) + \frac{\dot{y}(0)}{\Omega} \sin(\Omega t).
  \label{eq:HO_solution}
\end{equation}
From the relation Eq.~(\ref{eq:x_vs_y}), one finds
\begin{equation}
  y(0) = x(0), \ \ \ \ \ \ \ \dot{y}(0) = \dot{x}(0) + \frac{\gamma}{2} x(0).
  \label{eq:x_vs_y_initial}
\end{equation}
Substituting Eqs.~(\ref{eq:HO_solution}) and (\ref{eq:x_vs_y_initial}) into Eq.~(\ref{eq:x_vs_y}) leads to the solution of the damped harmonic oscillator Eq.~(\ref{eq:DHO_solution}).

In the above discussion, we observe that the transformation Eq.~(\ref{eq:x_vs_y}) replaces the damping term $\gamma \dot{x}(t)$ with the $ -(\gamma^2/4) y(t)$ term.
This replacement remains true even in the presence of an arbitrary external potential: 
suppose that $x(t)$ describes a dissipative motion with an arbitrary external potential $V(x,t)$, that is, $x(t)$ satisfies the following equation of motion,
\begin{equation}
  \ddot{x}(t) + \gamma \dot{x}(t) + \frac{1}{m} \frac{d V}{d x}(x(t),t) = 0.
  \label{eq:x_eom}
\end{equation}
By the transformation Eq.~(\ref{eq:x_vs_y}), the equation of motion for $y(t)$ is given as
\begin{equation}
\label{eq:y_eom}  \ddot{y}(t) + \frac{1}{m} e^{\gamma t / 2} \frac{d V}{d z}(z = y(t) e^{- \gamma t / 2},t) - \frac{\gamma^2}{4} y(t) = 0.
\end{equation}
This equation of motion is derived from the following Lagrangian,
\begin{equation}
  L_y (y, \dot{y}, t) = \frac{m}{2} \dot{y}^2 - \left[ e^{\gamma t} V(y e^{- \gamma t / 2},t) - \frac{m \gamma^2}{8} y^2 \right].
\end{equation}
This means that the transformation Eq.~(\ref{eq:x_vs_y}) maps a dissipative system to a system with the rescaled potential $e^{\gamma t} V(y e^{- \gamma t / 2},t)$ together with the inverted harmonic potential $- m \gamma^2 y^2 / 8$.
The damped harmonic oscillator discussed above is a 
special example where the potential is given by $V(x,t) = m \omega^2 x^2 / 2$ and the time-dependence of the rescaled potential disappears because of $\displaystyle e^{\gamma t} V(y e^{- \gamma t / 2}) = \frac{m \omega^2}{2} y^2$.

With the transformation Eq.~(\ref{eq:x_vs_y}), one can find an effective Lagrangian for a dissipative motion described by Eq.~(\ref{eq:x_eom}):
\begin{equation}
  L_x (x, \dot{x}, t) = L_y(y(x), \dot{y}(x,\dot{x}), t) = e^{\gamma t} \left[\frac{m}{2} \dot{x}^2 - V(x,t) \right] + \frac{m \gamma}{4} \frac{d}{dt} \left( x e^{\gamma t / 2} \right)^2.
\end{equation}
Neglecting the last term which does not affect the equation of motion, one obtains the Lagrangian of the Caldirola-Kanai model under an arbitrary time-dependent external potential
\begin{equation}
  L_{\rm CK} (x, \dot{x}, t) = e^{\gamma t} \left[\frac{m}{2} \dot{x}^2 - V(x,t) \right].
  \label{eq:L_CK_single}
\end{equation}

From these Lagrangians, we can derive the Hamiltonians for the motions of $x(t)$ and $y(t)$.
For a dissipative system described by the Caldirola-Kanai model, $x(t)$, it should be noted that the canonical momentum, $\pi_x (t)$, is different from the kinetic momentum, $m \dot{x}(t)$,
\begin{equation}
  \pi_x(t) = \frac{\partial L_{\rm CK}}{\partial \dot{x}} = m \dot{x}(t) e^{\gamma t}.
\end{equation}
One then obtains the Caldirola-Kanai Hamiltonian
\begin{equation}
  H_{\rm CK} (x, \pi_x, t) = \frac{\pi_x^2}{2m} e^{- \gamma t} + V(x,t) e^{\gamma t}.
  \label{eq:H_CK}
\end{equation}
For a system described by $y(t)$, on the other hand, the canonical momentum is equal to the kinetic momentum,
\begin{equation}
  \pi_y(t) = \frac{\partial L_y}{\partial \dot{y}} = m \dot{y}(t),
\end{equation}
and the Hamiltonian is given by
\begin{equation}
  H_{y} (y, \pi_y, t) = \frac{\pi_y^2}{2m} + e^{\gamma t} V(y e^{- \gamma t / 2},t) - \frac{m \gamma^2}{8} y^2.
  \label{eq:H_y}
\end{equation}

We note that the essence of this mapping between the dissipative motion of $x(t)$ and the dissipationless motion of $y(t)$ is the scale transformation Eq.~(\ref{eq:x_vs_y}). The slowing-down effect of the damping $\gamma$ is captured by the $e^{\frac{\gamma t}{2}}$ scale factor when relating the dissipative motion of $x(t)$ and dissipationless motion of $y(t)$~\cite{Kerner58,BJM94,HuangWu98}.



\subsection{\label{sec:Nonint_qm} Mapping in quantum mechanics}

Let us extend the mapping discussed in the previous subsection to quantum mechanics~\cite{Kerner58,BJM94,HuangWu98}.
From the Hamiltonians Eqs.~(\ref{eq:H_CK}) and (\ref{eq:H_y}), one can derive the corresponding Schr\"odinger equations.
For the Caldirola-Kanai model, Eq.~(\ref{eq:H_CK}), the Schr\"odinger equation reads
\begin{equation}
  i \hbar \frac{\partial}{\partial t} \phi(x,t) = \left[ - \frac{\hbar^2}{2m} e^{-\gamma t} \frac{\partial^2}{\partial x^2} + e^{\gamma t} V(x,t) \right] \phi(x,t),
  \label{eq:x_scheq_1}
\end{equation}
while for the other Hamiltonian without dissipation, one obtains
\begin{equation}
  i \hbar \frac{\partial}{\partial t} \psi(x,t) =  \left[ - \frac{\hbar^2}{2m} \frac{\partial^2}{\partial x^2}+ e^{\gamma t} V(x e^{-\gamma t/2},t) - \frac{m \gamma^2}{8} x^2 \right] \psi(x,t).
  \label{eq:y_scheq_1}
\end{equation}
The Schr\"odinger equation Eq.~(\ref{eq:x_scheq_1}) for $\phi(x,t)$ serves as a phenomenological modeling of quantum dissipative systems~\cite{Caldirola41,Kanai48}. Indeed, one can show that the Heisenberg equation of motion with Eq.~(\ref{eq:x_scheq_1}) has the same form as Eq.~(\ref{eq:x_eom}).
The equivalence of Eq.~(\ref{eq:x_scheq_1}) and Eq.~(\ref{eq:y_scheq_1}) can be understood clearly by showing that the wavefunctions $\phi$ and $\psi$ are related by a scale transformation as follows: we recall that the two models' solutions in the classical cases can be mapped to each other by the following transformation
\begin{equation}
    x(t) = y(t) e^{- \gamma t / 2}, \ \ \ \ \ \ \ \pi_x(t) = \left( \pi_y(t) - \frac{m \gamma}{2} y(t) \right) e^{\gamma t / 2}.
\end{equation}
In quantum mechanics, these relations should be satisfied as operators, not merely as expectation values.
In other words, $\phi(x,t)$ and $\psi(x,t)$ should satisfy
\begin{equation}
  \int_{-\infty}^{\infty} dz \, \phi^*(z,t) z^n \phi(z,t) = \int_{-\infty}^{\infty} dz \, \psi^*(z,t) \left( z e^{- \gamma t / 2} \right)^n \psi(z,t),
\end{equation}
and
\begin{equation}
  \int_{-\infty}^{\infty} dz \, \phi^*(z,t) \left( \frac{\hbar}{i} \frac{\partial}{\partial z} \right)^n \phi(z,t) = \int_{-\infty}^{\infty} dz \, \psi^*(z,t) \left\{ \left( \frac{\hbar}{i} \frac{\partial}{\partial z} - \frac{m \gamma}{2} z \right) e^{\gamma t / 2} \right\}^n \psi(z,t),
\end{equation}
for any natural number $n$.
From these conditions, we can find a relation between $\phi(x,t)$ and $\psi(x,t)$ up to a time-dependent phase, which should be determined to be consistent with the Schr\"odinger equations Eqs.~(\ref{eq:x_scheq_1}) and (\ref{eq:y_scheq_1}).
This consideration leads to the following relation between $\phi(x,t)$ and $\psi(x,t)$, 
\begin{equation}
  \phi(x,t) = \exp{\left( -i \frac{m \gamma}{4 \hbar} e^{\gamma t} x^2 + \frac{\gamma t}{4} \right)} \psi(x e^{\gamma t /2},t).
  \label{eq:phi_vs_psi_1}
\end{equation}
Indeed, one can directly show that $\phi(x,t)$ defined by Eq.~(\ref{eq:phi_vs_psi_1}) satisfies the Schr\"odinger equation with the Caldirola-Kanai Hamiltonian Eq.~(\ref{eq:x_scheq_1}) if $\psi(x,t)$ is a solution of Eq.~(\ref{eq:y_scheq_1}).
Therefore, if the initial conditions are related by
\begin{equation}
  \phi(x,0) = \exp{\left( -i \frac{m \gamma}{4 \hbar}x^2 \right)} \psi(x,0),
\end{equation}
one can find $\phi(x,t)$ from $\psi(x,t)$ through the equivalence Eq.~(\ref{eq:phi_vs_psi_1}), and vice verse.

The mapping dictates that the dissipative and dissipationless models obey essentially the same quantum dynamics through Eq.~(\ref{eq:phi_vs_psi_1}) and thus physically equivalent, as pointed out in the previous literature~\cite{HuangWu98}. The effect of the dissipation in Eq.~(\ref{eq:x_scheq_1}) is represented as a scale transformation between $x(t)$ and $y(t)$, and as the inverted harmonic barrier $- m \gamma^2  x^2/8$. In particular, the time-dependent scale transformation of the coordinate captures the slowing-down effect of the damping $\gamma$. This can be understood more directly by taking $V=0$: the dissipative quantum motion in free space of Eqs.~(\ref{eq:x_eom}) and~(\ref{eq:x_scheq_1}) are equivalent to the dissipationless quantum motion under an inverted harmonic barrier $- m \gamma^2  x^2/8$ in Eqs.~(\ref{eq:y_eom}) and~(\ref{eq:y_scheq_1}). Another important example is the harmonic oscillator potential $V(x,t) = m \omega^2 x^2 / 2$, which was discussed in the previous subsection. In this case, Eq.~(\ref{eq:y_scheq_1}) reads
\begin{equation}
  i \hbar \frac{\partial}{\partial t} \psi(x,t) = \left[ - \frac{\hbar^2}{2m} \frac{\partial^2}{\partial x^2} + \frac{m \Omega^2}{2} x^2 \right] \psi(x,t).
\end{equation}
On the other hand, the corresponding Caldirola-Kanai model describes a quantum damped harmonic oscillator.
Therefore, one can map the quantum dynamics of a simple harmonic oscillator to that of a damped harmonic oscillator, as was found in Refs.~\cite{Kerner58,BJM94,HuangWu98}.




\subsection{\label{sec:Nonint_mp} Extension to many-particle systems}

Let us extend the results in the previous subsections and in Refs.~\cite{Kerner58,BJM94,HuangWu98} for a single particle to an $N$-particle system in $d$-spacial dimensions in the presence of inter-particle interactions.
Our starting point is the dissipative equation of motion given by Eq.~(\ref{eq:Kanai}).
Instead of Eq.~(\ref{eq:x_vs_y}), we now consider the transformation ${\bf x}_i (t) = e^{- \gamma t / 2} {\bf y}_i (t)$ for $i = 1,\dots,N$ (note that all particles feel the same damping force, see Eq.~(\ref{eq:Kanai})).
As we derive Eq.~(\ref{eq:L_CK_single}) in Sec.\ref{sec:Nonint_cl}, the transformation leads to the Lagrangian given by Eq.~(\ref{eq:L_CK}).
Therefore, the Schr\"odinger equation for the dissipative motion reads
\begin{equation}
  i \hbar \frac{\partial}{\partial t} \phi({\bf x}_1, \dots, {\bf x}_N,t) = \left[ -  e^{-\gamma t} \sum_{i=1}^N \frac{\hbar^2 \nabla^2_{{\bf x}_i}}{2m_i} + e^{\gamma t} U({\bf x}_1, \dots, {\bf x}_N) \right] \phi({\bf x}_1, \dots, {\bf x}_N,t).
  \label{eq:x_scheq}
\end{equation}
On the other hand, the Schr\"odinger equation corresponding to Eq.~(\ref{eq:y_scheq_1}) reads
\begin{equation}
  i \hbar \frac{\partial}{\partial t} \psi({\bf x}_1, \dots, {\bf x}_N,t) = \left[ -  \sum_{i=1}^N \frac{\hbar^2 \nabla^2_{{\bf x}_i}}{2m_i}+ e^{\gamma t} U({\bf x}_1e^{-\gamma t/2}, \dots, {\bf x}_N e^{-\gamma t/2},t) - \sum_{i=1}^N \frac{m_i \gamma^2}{8} {\bf x}_i^2 \right] \psi({\bf x}_1, \dots, {\bf x}_N,t).
  \label{eq:y_scheq}
\end{equation}
As in the derivation of Eq.~(\ref{eq:phi_vs_psi_1}), we can find the relation between $\phi$ and $\psi$,
\begin{equation}
  \phi({\bf x}_1, \dots, {\bf x}_N,t) = \exp{\left( -i  e^{\gamma t} \sum_{i=1}^N \frac{m_i \gamma}{4 \hbar}{\bf x}_i^2 + \frac{d N \gamma t}{4} \right)} \psi({\bf x}_1e^{\gamma t/2}, \dots, {\bf x}_N e^{\gamma t/2},t).
  \label{eq:phi_vs_psi}
\end{equation}
This mapping for many-body quantum systems is a natural extension of what was studied for a single-particle motion in the previous literatures~\cite{Kerner58,BJM94,HuangWu98}.
The single-particle case in one dimension in Eq.~(\ref{eq:phi_vs_psi_1}) is indeed contained as $N=1$, $d=1$. When $U({\bf x}_1, \dots, {\bf x}_N,t) = \sum_{i=1}^N V({\bf x}_i,t)$ and hence each particle moves independently, we can regard them as a collection of non-interacting particles, each obeying the Schr\"odinger equations Eqs.~(\ref{eq:x_scheq_1}) and (\ref{eq:y_scheq_1}). When the potential $U$ cannot be represented as such a sum of one-body potentials, on the other hand, Eq.~(\ref{eq:x_scheq}) describes interacting $N$ quantum particles moving under 
the dissipation, while Eq.~(\ref{eq:y_scheq}) describes dissipationless motions of interacting quantum particles in the presence of the inverted harmonic barrier $- m \gamma^2  x_i^2/8$.



\section{Application to strongly interacting quantum systems}\label{sec:map_int}
We show in this section that the exact mapping formulated in the previous section 
can be utilized for a physical problem of a strongly interacting quantum system, namely the unitary Fermi gas~\cite{bertsch2001many,zwerger2011bcs,RevModPhys.80.1215,PhysRevLett.93.050401,nascimbene2010exploring,horikoshi2010measurement,ku2012revealing,gaebler2010observation,PhysRevLett.122.203401,levinsen2017universality}. The unitary Fermi gas is a system of spin-$1/2$ fermions interacting with an infinitely large $s$-wave scattering length. It has been recently realized with ultracold atoms~\cite{PhysRevLett.93.050401,nascimbene2010exploring,horikoshi2010measurement,ku2012revealing,gaebler2010observation,PhysRevLett.122.203401}, and it has been extensively studied because it is important for understanding nuclear matters and neutron star physics~\cite{bertsch2001many,zwerger2011bcs,RevModPhys.80.1215,horikoshi2019cold,PhysRevA.97.013601,endovirialsp,PhysRevC.82.045802}. The exact mapping between the dissipative and dissipationless motions should therefore be useful for understanding not only the dynamics of the unitary Fermi gas in ultracold atoms, but also for nuclear phenomena.

In Sec.~\ref{sec:Int_ZRInt}, we explain the interaction between the particles to model the unitary two-body interaction. In Sec.~\ref{sec:Int_proof}, we show that the exact mapping holds for a quantum system with the unitary interaction. In Sec.~\ref{sec:Int_dimstat}, we discuss the effects of dimensions and statistics of the particles. 
We argue that our exact mapping is non-trivial and useful for a three-dimensional system of fermions, while it is either trivial or breaks down for low-dimensional systems and for 
bosonic systems.

\subsection{\label{sec:Int_ZRInt}Zero-range interaction}
Solutions of interacting quantum systems in general sensitively depends on details of the interaction potentials. However, for low-energy quantum systems, details of the interaction become irrelevant, so that any Hamiltonian can be universally described by a so-called pseudo-potential. This universality originates from the fact that the scattering between the particles at low energy are dominated by the $s$-wave scattering, which is universally characterized solely by the $s$-wave scattering length $a$ for non-relativistic low-energy quantum systems interacting via short-range interactions. As long as this condition holds, the interaction potential between the $i$-th and $j$-th particles $V_{ij}$ can be legitimately replaced by the following pseudo potential introduced by Huang and Yang in three spacial dimension~\cite{PhysRev.105.767}
\begin{equation}
\label{eq:HYpp}V_{ij}({\bf r}_{ij})= \frac{2\pi a_{ij}\hbar^2}{\mu_{ij}}\delta^{(3)}({\bf r}_{ij})\frac{\partial}{\partial r_{ij}}r_{ij},
\end{equation}
where $r_{ij}=|{\bf r}_{i}-{\bf r}_{j}|$, and $a_{ij}$ and $\mu_{ij}=m_i m_j/(m_i +m_j)$ are the $s$-wave scattering length and the reduced mass of the $i$-th and $j$-th particles, respectively. The Hamiltonian can then be wrriten as the sum of the independent-particle Hamiltonian $H_0$ and the interaction term $U$, $H = H_0 + U$ with
\begin{equation}
U({\bf x}_1, \dots, {\bf x}_N,t) =  \sum_{i<j} V_{ij}({\bf r}_{ij}).
\end{equation}
An alternative low-energy universal approach can be found by noting that the solution of the Huang and Yang pseudo-potential has the following asymptotic form when the particles get close together
\begin{equation}
\label{eq:BPbc}\lim_{r_{ij}\rightarrow 0} \Psi({\bf r}_{1},{\bf r}_{2},...,{\bf r}_{N}) = \left(\frac{1}{r_{ij}}-\frac{1}{a_{ij}}\right) A({\bf R}_{ij},{\bf r}_{1},...,{\bf r}_{i-1},{\bf r}_{i+1}...,{\bf r}_{N}),
\end{equation}
where ${\bf R}_{ij} =(m_i {\bf r}_{i}+ m_j {\bf r}_{j})/(m_i +m_j) $ is the center of mass between the $i$-th and $j$-th particles, and $A$ is the regular part of the wavefunction $\Psi$ when the limit $r_{ij}\rightarrow 0$ is taken. The singular term $1/r_{ij}-1/a_{ij}$ represents the universal $s$-wave scattering behavior of the two-body scattering. Therefore, we can simply replace the effect of the interaction term $V_{ij}$ with this boundary condition: we solve the free particles' equation of motion of $H = H_0$ supplemented with the boundary condition Eq.~(\ref{eq:BPbc})

The above two approaches, the Huang-Yang pseudo-potential Eq.~(\ref{eq:HYpp}) and the Bethe-Peierls boundary condition Eq.~(\ref{eq:BPbc}), are equivalent, and have both been widely used for low-energy dilute systems where $s$-wave interaction is predominantly large. In particular, they have been successfully used to study ultracold atoms close to the unitary limit $|a_{ij}| \rightarrow \infty$~\cite{zwerger2011bcs,RevModPhys.80.1215,PhysRevA.97.013601,endovirialsp,nielsen2001three,naidon2017efimov}. It has also been used in low-energy nuclear physics as the $s$-wave scattering length between the nucleons are very large~\cite{bertsch2001many,horikoshi2019cold,PhysRevC.82.045802}. We note however that Eq.~(\ref{eq:HYpp}) and Eq.~(\ref{eq:BPbc}) are only valid in three spacial dimension: they need to be modified in the other spacial dimension, as will be discussed in Sec.~\ref{sec:Int_dimstat}. We also note that it is required that three- and higher-body interactions are negligibly small to legitimately use the pseudo-potential and the boundary condition methods, as will also be discussed in more details in Sec.~\ref{sec:Int_dimstat}.

\subsection{Exact mapping for the unitary interacting system}\label{sec:Int_proof}

Let us first consider $N$-particle system where the inter-particle interactions are modeled by the Bethe-Peierls boundary condition Eq.~(\ref{eq:BPbc}). As the boundary condition method deals with the non-interacting Hamiltonian $U=0$, the Schr\"odinger equations for the Caldirola-Kanai model and the corresponding dissipationless model are the same as Eqs.~(\ref{eq:x_scheq}) and (\ref{eq:y_scheq}) with $U=0$.
Thus, we can naturally expect that the mapping Eq.~(\ref{eq:phi_vs_psi}) should also relate these two problems. This turns out to be true if the $s$-wave scattering length of the two systems satisfy a certain relation (see Eq.~(\ref{eq:a_relation})).

To show this, let us consider the Bethe-Peierls boundary condition for the Caldirola-Kanai model under the unitary interaction
\begin{equation}
\label{eq:asymp_x}\phi({\bf x}_{1},{\bf x}_{2},...,{\bf x}_{N},t) \propto \left(\frac{1}{x_{ij}}-\frac{1}{a_{ij}^{(\mathrm{CK})}(t)}\right).
\end{equation}
Here $a_{ij}^{(\mathrm{CK})}(t)$ is the $s$-wave scattering length between the $ij$ particles, 
which is now allowed to vary with time for later purposes.
Note that the Caldirola-Kanai model with $U=0$ describes the motion of non-interacting particles under the influence of a damping force (see Eq.~(\ref{eq:Kanai}) with $U=0$). With the boundary condition Eq.~(\ref{eq:asymp_x}), those damped particles interact through the $s$-wave scattering length $a_{ij}^{(\mathrm{CK})}(t)$ once two particles come to the same position.
Assuming that the mapping Eq.~(\ref{eq:phi_vs_psi}) holds and noting that the exponential phase factor does not affect the singular short-distance behavior of the wavefunction, we obtain the following boundary condition of the corresponding dissipationless system
\begin{equation}
\label{eq:asymp_y}\psi({\bf y}_{1},{\bf y}_{2},...,{\bf y}_{N},t) \propto \left(\frac{1}{y_{ij}}-\frac{1}{e^{\frac{\gamma t}{2}}a_{ij}^{(\mathrm{CK})}(t)}\right).
\end{equation}
Therefore, we find that the $s$-wave scattering length of the $\psi$ system, $a_{ij}^{(\mathrm{H})}(t)$, must be given by
\begin{equation}
\label{eq:a_relation}a_{ij}^{(\mathrm{H})}(t)=e^{\frac{\gamma t}{2}}a_{ij}^{(\mathrm{CK})}(t).
\end{equation}

While we have used above the Bethe-Peierls boundary condition to derive the condition Eq.~(\ref{eq:a_relation}) for the equivalence of the two systems, we can alternatively derive it with the Huang-Yang pseudo-potential Eq.~(\ref{eq:HYpp}). Indeed, Eqs.~(\ref{eq:x_scheq}) and (\ref{eq:y_scheq}) with the Huang-Yang interaction read
\begin{equation}
  i \hbar \frac{\partial}{\partial t} \phi({\bf x}_1, \dots, {\bf x}_N,t) = \Biggl[ -  e^{-\gamma t} \sum_{i=1}^N \frac{\hbar^2\nabla^2_{{\bf x}_i}}{2m_i} + e^{\gamma t}\sum_{i<j} \frac{2\pi a_{ij}(t)\hbar^2}{\mu_{ij}}\delta^{(3)}({\bf x}_{ij})\frac{\partial}{\partial x_{ij}}x_{ij} \Biggr] \phi({\bf x}_1, \dots, {\bf x}_N,t).
  \label{eq:x_scheq2}
\end{equation}
\begin{equation}
  i \hbar \frac{\partial}{\partial t} \psi({\bf x}_1, \dots, {\bf x}_N,t) = \Biggl[ -  \sum_{i=1}^N \frac{\hbar^2\nabla^2_{{\bf x}_i}}{2m_i} + e^{\frac{5\gamma}{2} t}\sum_{i<j} \frac{2\pi a_{ij}(t)\hbar^2}{\mu_{ij}}\delta^{(3)}({\bf x}_{ij})\frac{\partial}{\partial x_{ij}}x_{ij} - \sum_{i=1}^N \frac{m_i \gamma^2 x_i^2}{8} \Biggr] \psi({\bf x}_1, \dots, {\bf x}_N,t).
  \label{eq:y_scheq2}
\end{equation}
We then need to relate $a_{ij}(t)$ with the $s$-wave scattering lengths of the two systems $a_{ij}^{(\mathrm{CK})}(t)$, $a_{ij}^{(\mathrm{H})}(t)$. This can be done by substituting the Eq.~(\ref{eq:asymp_x}) and Eq.~(\ref{eq:asymp_y}) to the right-hand sides of Eq.~(\ref{eq:x_scheq2}) and Eq.~(\ref{eq:y_scheq2}), respectively. We then find
\begin{equation}
a_{ij}^{(\mathrm{CK})}(t) = e^{2\gamma t}a_{ij}(t),  \ \ \ \ a_{ij}^{(\mathrm{H})}(t) = e^{\frac{5}{2}\gamma t}a_{ij}(t),
\end{equation}
from which we arrive at Eq.~(\ref{eq:a_relation}).


As the condition Eq.~(\ref{eq:a_relation}) must be satisfied for all pairs of particles $ij$ at any time, our exact mapping is difficult to hold true in general. However, there are two special cases where Eq.~(\ref{eq:a_relation}) is easily satisfied. The first case is the non-interacting system $a_{ij}^{(\mathrm{H})}=a_{ij}^{(\mathrm{CK})} =0$. This corresponds to the system considered in the previous section and in Refs~\cite{Kerner58,BJM94,HuangWu98}. The other more interesting case is the unitary gas $a_{ij}^{(\mathrm{H})} =a_{ij}^{(\mathrm{CK})} = \pm\infty $. We thus arrive at the following non-trivial conclusion for the two strongly interacting quantum dynamics: a dissipative quantum motion of the unitary gas described by Eq.~(\ref{eq:x_scheq2}) is equivalent to the dissipationless quantum dynamics of the unitary gas under an inverted harmonic potential in Eq.~(\ref{eq:y_scheq2}) via the mapping Eq.~(\ref{eq:phi_vs_psi}).

The physical reason for the two seemingly opposite conditions, the non-interacting system and the unitary system, can be ascribed to the scale invariance. The mapping between the Caldirola-Kanai system and the dissipationless inverted harmonic system can be regarded as a sort of scale transformation in Eq.~(\ref{eq:x_vs_y}). With this transformation, the interaction term is transformed as  (see Eqs.~(\ref{eq:x_scheq}) and~(\ref{eq:y_scheq}))
\begin{equation}
U({\bf x}_1, \dots, {\bf x}_N) \rightarrow e^{\gamma t}U({\bf x}_1e^{-\gamma t/2}, \dots, {\bf x}_N e^{-\gamma t/2},t).
\end{equation}
In general, they show rather different Hamiltonian dynamics. However, if the interaction term satisfies the scale invariance
\begin{equation}
e^{\gamma t} U({\bf x}_1e^{-\gamma t/2}, \dots, {\bf x}_N e^{-\gamma t/2},t) = f(t)U({\bf x}_1, \dots, {\bf x}_N ,t)
\end{equation}
 with $ f(t)$ an arbitrary time-dependent function, the Hamiltonian keeps its form with this scale transformation, showing essentially the same dynamics. Indeed, the non-interacting system and the unitary system are known to be scale invariant because the length scale characterizing the interaction, the $s$-wave scattering length, disappears.  In addition to the zero-range type interaction, we can also consider more general classes of scale-invariant interactions as will be discussed in the Appendix A. We can indeed show that the exact mapping holds true for various classes of scale-invariant interactions as long as the interaction strength satisfies a similar relation as Eq.~(\ref{eq:a_relation}) (see Appendix A).  We also note that our mapping for the unitary system can be alternatively proved using the scaling solution of the unitary Fermi gas in the time-dependent harmonic trap~\cite{PhysRevA.86.013626,PhysRevA.74.053604} (see Appendix B).


\subsection{\label{sec:Int_dimstat}Effects of dimensions, quantum statistics of the particles, and higher-body interactions}
We note that the Huang and Yang pseudo-potential Eq.~(\ref{eq:HYpp}) and the Bethe-Peierls boundary condition Eq.~(\ref{eq:BPbc}) are only valid in three spacial dimensions. For low-dimensional systems, the Bethe-Peierls boundary condition should be modified as~\cite{RevModPhys.80.1215,naidon2017efimov,PhysRevLett.81.938,PhysRevA.86.013626}
\begin{equation}
\label{eq:BPbc_2D}\lim_{r_{ij}\rightarrow 0} \Psi({\bf r}_{1},{\bf r}_{2},...,{\bf r}_{N}) = \ln\left(\frac{r_{ij}}{a_{ij}^{(\mathrm{2D})}}\right) A({\bf R}_{ij},{\bf r}_{1},...,{\bf r}_{i-1},{\bf r}_{i+1}...,{\bf r}_{N}),
\end{equation}
for a two-dimensional system, and
\begin{equation}
\label{eq:BPbc_1D}\lim_{r_{ij}\rightarrow 0} \Psi({\bf r}_{1},{\bf r}_{2},...,{\bf r}_{N})= \\
\left(|x_{ij}|- a_{ij}^{(1D)}\right) A({\bf R}_{ij},{\bf r}_{1},...,{\bf r}_{i-1},{\bf r}_{i+1}...,{\bf r}_{N})
\end{equation}
for a one-dimensional system, respectively. Here, $a_{ij}^{(\mathrm{2D})}$ and $a_{ij}^{(\mathrm{1D})}$ are two-dimensional and one-dimensional scattering lengths, respectively. By using these boundary conditions, we can follow almost the same argument as in Sec~\ref{sec:Int_proof}. We then arrive at the same conclusion: the two-dimensional and one-dimensional systems of the Caldirola-Kanai model Eq.~(\ref{eq:x_scheq}) and the corresponding inverted harmonic model Eq.~(\ref{eq:y_scheq}) are equivalent via the mapping Eq.~(\ref{eq:phi_vs_psi}) if and only if the $s$-wave scattering lengths of the two systems satisfy Eq.~(\ref{eq:a_relation}) for any $ij$ particles at any time. This can be also shown by using the Huang-Yang pseudo potential in two and one dimension~\cite{PhysRevLett.88.010402,rontani2017renormalization}
\begin{equation}
\label{eq:HYpp_2D}V_{ij}({\bf r}_{ij})=- \frac{\pi \hbar^2 \delta^{(2)}({\bf r}_{ij})}{\mu_{ij}\ln(q \Lambda a_{ij}^{(\mathrm{2D})})} w\left[1- \ln(q \Lambda r_{ij})r_{ij}\frac{\partial}{\partial r_{ij}} \right]  \ \ \ (\mathrm{2D}),
\end{equation}
\begin{equation}
\label{eq:HYpp_1D}V_{ij}(x_{ij})=- \frac{\hbar^2}{\mu_{ij}a_{ij}^{(\mathrm{1D})}}\delta(x_{ij}) \ \ \ (\mathrm{1D}),
\end{equation}
where $q=\frac{1}{2}e^{C}$ with $ C=0.577...$ is the Euler gamma, and $\Lambda$ is an arbitrary momentum scale.

It is then tempting to conclude that our exact mapping is also valid for the strongly interacting two-dimensional and one-dimensional systems. However, this is not the case: our mapping for the unitary interacting system should be non-trivial only for three-dimensional system. Indeed, the unitary system $a=\pm \infty$ is known to be non-interacting and thus trivial in two and one dimensions, which is in stark contrast to the three dimensional system where it is genuine strongly interacting system~\cite{zwerger2011bcs,RevModPhys.80.1215,PhysRevA.86.013626,PhysRevA.86.053633}. This well-known fact can be easily understood from Eqs.~(\ref{eq:HYpp_2D}) and~(\ref{eq:HYpp_1D}), where the interaction becomes non-interacting $V_{ij} =0$ when $a=\pm\infty$. This is in contrast to Eq.~(\ref{eq:HYpp}) in three dimension where $V_{ij}\neq 0$ when $a=\pm\infty$. Therefore, our mapping for low dimensional quantum systems is trivial and has limited range of applications.

We note that we have solely considered two-body interaction, neglecting three- and higher-body interactions.
In most interacting quantum systems, the two-body interaction is much more relevant than the three- and higher-body interactions, even when it is present. Thus, our assumption seems plausible. We should remark however that the unitary interacting system may essentially require the three- and higher-body interactions. A prime example is a system of identical bosons interacting via the unitary interaction $a=\pm\infty$, where the Efimov effect occurs~\cite{nielsen2001three,naidon2017efimov,efimov1970energy,kraemer2006evidence}. When the Efimov effect occurs, the Hamiltonian with only the unitary two-body interaction in Eq.~(\ref{eq:HYpp}) becomes singular, and one needs to introduce the three-body boundary condition to make the Hamiltonian well-defined~\cite{PhysRev.47.903,nielsen2001three,naidon2017efimov,PhysRevLett.112.105301}. As this three-body boundary condition is generally not scale invariant, our exact mapping breaks down. The Efimov effect generally occurs for three-dimensional unitary interacting bosonic systems, and thus our result is not applicable. 
On the other hand, the Efimov effect does not tend to occur for fermionic system due 
to the Pauli exclusion principle. In particular, it is shown that the spin-1/2 identical fermions with the unitary interaction Eq.~(\ref{eq:HYpp}), i.e. the unitary Fermi gas, is well-defined as there is no three-body nor higher-body Efimov effect~\cite{efimov1973energy,PhysRevA.67.010703,endo2011universal,endo2012universal,PhysRevA.92.053624,PhysRevLett.118.083002,naidon2017efimov}. Our result should therefore be useful for the unitary Fermi gas. Furthermore, as the mass-imbalanced Fermi system does not show the Efimov effect either when the mass imbalance is $M/m \lesssim 13$~\cite{naidon2017efimov,efimov1973energy,PhysRevA.67.010703,PhysRevLett.118.083002,PhysRevLett.105.223201}, we can state that we can readily apply our exact mapping to such mass-imbalanced two-component unitary Fermi gas systems~\cite{PhysRevLett.112.075302,PhysRevA.93.053611}.

We also note that the Efimov effect does not occur for two-dimensional and one-dimensional systems~\cite{PhysRevA.86.013626,PhysRevA.86.053633,PhysRevA.19.425,nielsen2001three,naidon2017efimov}. The Hamiltonian with the two-body interaction Eqs.~(\ref{eq:HYpp_2D}) and~(\ref{eq:HYpp_1D}) are thus well-defined even without the higher-body interaction, and our argument on the two- and one-dimensional systems above are valid regardless of the quantum statistics of the particles. We repeat however that the two-dimensional and one-dimensional unitary systems correspond to non-interacting systems, so that the mapping would not be so useful.



\section{Conclusion and Discussions}\label{sec:concl}
We have studied a quantum dissipative dynamics of the Caldirola-Kanai model (see Eqs.~(\ref{eq:Kanai}) and~(\ref{eq:x_scheq})), 
and shown that it can be rigorously mapped to a dissipationless quantum dynamics under an inverted-harmonic potential (see Eq.~(\ref{eq:y_scheq})). 
While this mapping has been known for a single-particle and non-interacting systems~\cite{Kerner58,BJM94,HuangWu98}, we have shown that it also holds true for strongly interacting systems. In particular, we have found that the dissipative dynamics of the unitary Fermi gas can be exactly mapped to a dissipationless dynamics of the unitary Fermi gas under an inverted harmonic potential.

The unitary Fermi gas has been recently realized in cold atom experiments~\cite{PhysRevLett.93.050401,nascimbene2010exploring,horikoshi2010measurement,ku2012revealing,gaebler2010observation,PhysRevLett.122.203401,bardon2014transverse,cao2011universal}. It has attracted a lot of research interests, because unveiling the properties of the unitary Fermi gas is important for understanding nuclear matters and neutron star physics~\cite{bertsch2001many,zwerger2011bcs,RevModPhys.80.1215,horikoshi2019cold,PhysRevA.97.013601,endovirialsp,PhysRevC.82.045802}. Our work therefore should be particularly useful for cold atoms and nuclear systems. The external potential can be well-controlled in cold atom experiments with laser and magnetic field, and it is possible to engineer the harmonic or inverted-harmonic potential~\cite{zwerger2011bcs,RevModPhys.80.1215} and observe its quantum dynamics. It is thus a promising system where our mapping would be useful. We also note that we can control the $s$-wave scattering length in a time-dependent manner in cold atoms~\cite{clark2017collective}. It is thus possible to realize a situation where our condition in Eq.~(\ref{eq:a_relation}) holds, so that our exact mapping is valid for systems with finite scattering length or low-dimensional systems.

A more challenging but interesting system to apply our result is nuclear physics. Nuclear systems composed of protons and neutrons can be approximately regarded as the unitary Fermi system as the $s$-wave scattering lengths between the nucleons are large. With recent studies on nuclear reactions, it has been pointed out that the dissipative quantum tunneling may play an important role in low-energy fusion reactions~\cite{Dasgupta07,hagino2012subbarrier,Rafferty16}. It is also suggested that quantum dissipations of the nuclear matters are indispensable for understanding the dynamics of the neutron stars~\cite{cutler1987effect,PhysRevLett.120.041101}. Although we should remark that the Caldirola-Kanai model is a rather simple toy model and that the range corrections and three-body interaction neglected in our study often turns out to be relevant in the nuclear systems~\cite{PhysRevC.66.064001,RevModPhys.85.197,PhysRevLett.112.105301}, our analytical mapping between the dissipative and dissipationless systems can provide us with novel qualitative perspective to understand those nuclear phenomena.

\section*{Appendix A: Exact mapping for scale-invariant interactions}\label{sec:app1}
As noted in Sec.~\ref{sec:Int_proof}, the exact mapping between the Caldirola-Kanai model and the dissipationless inverted harmonic potential model holds not only for the zero-range interaction, but also for various classes of scale-invariant interactions. To see this point, let us first consider a power-law two-body interaction $U({\bf x}_1, \dots, {\bf x}_N)= \sum_{i<j} \alpha_{ij}(t)x_{ij}^{\beta}$, where $x_{ij}= |{\bf x}_i-{\bf x}_j|$. From Eqs.~(\ref{eq:x_scheq2}) and~(\ref{eq:y_scheq2}), the Schr\"odinger equations for the Caldirola-Kanai model and the corresponding dissipationless harmonic model reads
\begin{equation}
  i \hbar \frac{\partial}{\partial t} \phi({\bf x}_1, \dots, {\bf x}_N,t) = \Biggl[ -  e^{-\gamma t} \sum_{i=1}^N \frac{\hbar^2 \nabla^2_{{\bf x}_i}}{2m_i} +  e^{\gamma t} \sum_{i<j} a_{ij}(t) x_{ij}^{\beta} \Biggr] \phi({\bf x}_1, \dots, {\bf x}_N,t).
  \label{eq:x_scheq_ap1}
\end{equation}
\begin{equation}
  i \hbar \frac{\partial}{\partial t} \psi({\bf x}_1, \dots, {\bf x}_N,t) = \Biggl[ -  \sum_{i=1}^N \frac{\hbar^2 \nabla^2_{{\bf x}_i}}{2m_i} +  e^{\gamma t}e^{-\frac{\beta \gamma}{2} t} \sum_{i<j} a_{ij}(t) x_{ij}^{\beta} - \sum_{i=1}^N \frac{m_i \gamma^2  x_i^2}{8} \Biggr] \psi({\bf x}_1, \dots, {\bf x}_N,t).
  \label{eq:y_scheq_ap1}
\end{equation}

As in Sec.~\ref{sec:Int_proof}, we can introduce the effective interaction coupling constants of the Caldirola-Kanai model $\alpha_{ij}^{(\mathrm{CK})}$ and the dissipationless harmonic model $\alpha_{ij}^{(\mathrm{H})}$ as follows
\begin{equation}
 \label{eq:a_relation_ap0}\alpha_{ij}^{(\mathrm{CK})}(t) = e^{2\gamma t}a_{ij}(t),  \ \ \ \ \alpha_{ij}^{(\mathrm{H})}(t) = e^{\gamma t}e^{-\frac{\beta \gamma}{2} t} a_{ij}(t).
\end{equation}
We thus arrive at the following condition:
\begin{equation}
\label{eq:a_relation_ap}\alpha_{ij}^{(\mathrm{H})}(t) = e^{-\gamma\left(1+\frac{\beta}{2}\right) t}\alpha_{ij}^{(\mathrm{CK})}(t).
\end{equation}
If this condition is satisfied at any time, the two models become equivalent by Eq.~(\ref{eq:phi_vs_psi}). The above argument is parallel to the zero-range interaction discussed in Sec.~\ref{sec:Int_proof}. Indeed, the zero-range interaction corresponds to the scale factor $\beta=-3$, and therefore Eq.~(\ref{eq:a_relation}) in Sec.~\ref{sec:Int_proof} can be regarded as the special case of Eq.~(\ref{eq:a_relation_ap}).

A particularly interesting case is the inverse power-law interaction $\beta=-2$. In this case, Eq.~(\ref{eq:a_relation_ap}) becomes $\alpha_{ij}^{(\mathrm{H})}(t) =\alpha_{ij}^{(\mathrm{CK})}(t)$, i.e. the two models are equivalent for the same interaction coupling strength. In particular, the two models are equivalent for a time-independent interaction strength $\alpha_{ij}^{(\mathrm{H})}=\alpha_{ij}^{(\mathrm{CK})}={\rm Const}$. This originates from the special symmetry of the quantum system with $r^{-2}$ interaction.

The above argument essentially relies on the scaling property of the power-law homogeneous interaction. We can further generalize the exact mapping for more exotic inhomogeneous interactions. For example,
\begin{equation}
U=  \alpha_{ij}(t)\sum_{i<j}\left[ \frac{1}{x_{ij}^3}+\delta^{(3)}({\bf x}_{ij})\frac{\partial}{\partial x_{ij}}x_{ij}\right]
\end{equation}
shows the same scaling property as the power-law interaction $x_{ij}^{-3}$, and therefore the exact mapping holds for Eq.~(\ref{eq:a_relation_ap}) with $\beta=-3$. The above interaction may be realized in dipoler cold atoms~\cite{PhysRevLett.94.160401,baranov2012condensed}: cold atoms with large magnetic dipole moments, where atoms interact via strong magnetic dipole-dipole interaction plus a short-range contact interaction. We can also consider three-body and higher-body interactions, such as $U({\bf x}_1, \dots, {\bf x}_N)=  \sum_{ijk}\alpha_{ijk}(t) x_{ij}^{\beta+a} x_{jk}^{b} x_{kl}^{-a-b} $ or $U({\bf x}_1, \dots, {\bf x}_N)=  \sum_{i<j}\alpha_{ijkl}(t)  x_{ij}^{\beta+a} x_{jk}^{b} x_{kl}^{c} x_{li}^{-a-b-c}$ where $a,b$, and $c$ are arbitrary constants. These seemingly complicated three-body and four-body interactions show essentially the same scaling property as the power-law two-body interaction discussed above. Therefore, our exact mapping holds true for these systems with higher-body interactions under the same condition Eq.~(\ref{eq:a_relation_ap}), including the special case of $\beta=-2$ with $\alpha_{ijk}, \alpha_{ijkl} = {\rm Const}$.

\section*{Appendix B: Exact mapping derived with the scaling solution of the unitary Fermi gas}\label{sec:app2}
We present in this appendix 
an alternative way to prove our exact mapping. We start from the scaling solution of the unitary Fermi gas~\cite{PhysRevA.74.053604}. When the unitary Fermi gas evolves in time under a time-dependent harmonic trap $\omega(t)$ for $t>0$ after held in a constant trapping frequency $\omega(t\le 0) =\omega_0$, its wave function at an arbitrary time $t>0$ can be found by evolving the $t=0$ solution
\begin{equation}
\label{eq:scal_sol_rel}\Psi({\bf x}_{1},...,{\bf x}_{N},t) = \frac{1}{\lambda(t)^{3N/2}}\exp\left[ i  \sum_{i=1}^N \frac{m_i \dot{\lambda}}{2\hbar \lambda}{\bf x}_i^2 \right] \tilde{\Psi}\left(\frac{{\bf x}_{1}}{\lambda(t)},...,\frac{{\bf x}_{N}}{\lambda(t)},\tau(t)\right),
\end{equation}
where $\Psi$ and $\tilde{\Psi}$ denote the wavefunction of the unitary Fermi gas under a time-dependent trap $\omega(t)$  and time-independent trap $\omega_0 $, respectively. The scale-factor $\lambda(t)$ and effective time $\tau(t)$ are defined as
\begin{equation}
\label{eq:dif_eq_lam}\ddot{\lambda}= \frac{\omega_0^2}{\lambda^3}- \omega^2(t) \lambda,
\end{equation}
\begin{equation}
\tau(t) = \int_0^{t}\frac{dt'}{\lambda^2(t')}.
\end{equation}

Let us consider the following time-dependent harmonic trap
\[
\omega(t) =
\left\{
\begin{array}{cl}
-\gamma^2/4 & (t\le 0), \\
 0& (t>0).
\end{array}
\right.
\]
Eq.~(\ref{eq:dif_eq_lam}) reads
\begin{equation}
\ddot{\lambda}= -\frac{\gamma^2}{4 \lambda^3} \ \ \ (t>0),
\end{equation}
from which we find  $\lambda(t)=\sqrt{1-\gamma t}$ and $\tau(t)=-1/\gamma \ln(1-\gamma t)$. As the wavefunction in the right-hand side of Eq.~(\ref{eq:scal_sol_rel}) corresponds to the dissipationless unitary gas wavefunction under a harmonic potential $-m\gamma^2 x_i^2/8$, we can regard $\tilde{\Psi}$ as $\psi$ in our exact mapping (see Eq~(\ref{eq:y_scheq}) with $U=0$). On the other hand, the wavefunction in the right-hand side of Eq.~(\ref{eq:scal_sol_rel}) corresponds to the free-space unitary Fermi gas which obeys the Schr\"odinger equation
\begin{equation}
  i \hbar \frac{\partial}{\partial t} \tilde{\Psi}({\bf x}_1, \dots, {\bf x}_N,t) =- \sum_{i=1}^N \frac{\hbar^2 \nabla^2_{{\bf x}_i}}{2m_i} \tilde{\Psi}({\bf x}_1, \dots, {\bf x}_N,t).
 \label{eq:x_scheq_appdif1}
\end{equation}
To relate this with the Caldirola-Kanai model, we invert the relation between $\tau$ and $t$ as $t(\tau)=(1-e^{-\gamma \tau})/\gamma$. We then find
\begin{equation}
  i \hbar \frac{\partial}{\partial \tau} \tilde{\Psi}({\bf x}_1, \dots, {\bf x}_N,t) = - e^{-\gamma \tau}\sum_{i=1}^N \frac{\hbar^2 \nabla^2_{{\bf x}_i}}{2m_i} \tilde{\Psi}({\bf x}_1, \dots, {\bf x}_N,t).
 \label{eq:x_scheq_appdif2}
\end{equation}
This equation is the same as the Caldirola-Kanai model in Eq~(\ref{eq:x_scheq}) with $U=0$, so we can regard $\tilde{\Psi}$ as $\phi$. The $\lambda$ and $\dot{\lambda}/\lambda$ in the global phase factor in Eq.~(\ref{eq:scal_sol_rel}) reads  $\lambda(t)=\sqrt{1-\gamma t}=e^{-\frac{\gamma \tau}{2}}$ and $\dot{\lambda}/\lambda = -\frac{\gamma}{2}e^{\gamma \tau}$, from which we obtain our exact mapping Eq.~(\ref{eq:phi_vs_psi}).


\begin{acknowledgments}
  We thank Kouchi Hagino, Yvan Castin, Hajime Togashi, and Yusuke Tanimura for fruitful discussions. In particular, some discussions in the Appendix B are inspired by private communication with Yvan Castin.
  This work was supported by Tohoku University Graduate Program on Physics for the Universe (GP-PU), and JSPS KAKENHI Grant Numbers JP18J20565 and 21H00116.
\end{acknowledgments}


\end{document}